# Towards an Intelligent Assistive System Based on Augmented Reality and Serious Games

Fatemeh Ghorbani[1], Mahsa Farshi Taghavi[1], Mehdi Delrobaei[1, 2]

[1] *Faculty of Electrical Engineering, K. N. Toosi University of Technology, Tehran 1631714191, Iran.*
[2] *Department of Electrical and Computer Engineering, Western University, London, ON N6A 5B9, Canada.*

*E-mail addresses:*
fatemeghorbani@email.kntu.ac.ir, m.farshitaghavi@email.kntu.ac.ir, delrobaei@kntu.ac.ir, https://orcid.org/my-orcid?orcid=0000-0002-4188-6958

*Abstract* - Age-related cognitive impairment is generally characterized by gradual memory loss and decision-making difficulties. The aim of this study is to investigate multi-level support and suggest relevant helping means for the elderly with mild cognitive impairment as well as their caregivers as the primary end-users. This work reports preliminary results on an intelligent assistive system, achieved through the integration of Internet of Things, augmented reality, and adaptive fuzzy decision-making methods. The proposed system operates in different modes, including automated and semi-automated modes. The former helps the user complete their daily life activities by showing augmented reality messages or making automatic changes; while the latter allows manual changes after the real-time assessment of the user's cognitive state based on the augmented reality serious game score. We have also evaluated the accuracy of the serious game score with 37 elderly participants and compared it with users' paper-based cognitive test results. We further noted that there is an acceptable correlation between the paper-based test and users' serious game scores. Moreover, we observed that the system response in the semi-automated mode causes less data loss compared with the automated mode, as the number of active devices decreases.

## 1. Introduction

Recent projections of the world population indicate obvious trends tending towards more elderly people with the proportion of the global population aged 65 years or over is expected to increase from 9.3 percent in 2020 to 16.0 percent by 2050 [1]. Moreover, age-related cognitive impairment is a degenerative neurological disorder charac- terized by progressive loss of memory affecting the elderly [2]. Loss of memory and lacking capacity to make decisions are two main difficulties experienced by the elderly and patients with cognitive impairments. Nearly 5.8 million patients are currently suffering from Alzheimer's disease (AD), as the common cause of cognitive impairments, in the United States alone [3]. They generally have problems in remembering recent information and completing everyday tasks. Therefore, they should always be reminded of the required tasks that improve their confidence and quality of life [4].

In response to the increasing number of elder people with mild cognitive impairment (MCI) and a lack of treatment that slows or stops its progression, the pervasive deployment of intelligent assistive (IA) systems could have a disruptive effect on the elderly and their care-
givers' daily life [5]. IA systems can (i) reduce the burden on public finances throughout the postponement or removal of institutional care,
(ii) lessen the psychological burden on formal caregivers and families,
(iii) compensate for the lack of human caregivers while improving and optimizing the quality of care, and (iv) empower older adults with MCI and thereby enhancing their confidence [6]. For example, in [7], the authors proposed *MED-AR* system based on the intelligent augmented reality (AR) for helping older adults in the medication task. The system was designed to present a research methodology for tracking and distributing prescribed medicines for older adults in a home health care scenario.

On the other hand, serious games have had positive impacts on reconstructing a functional environment where the user could poten- tially encounter real-world scenarios [8–10]. Moreover, researchers have shown that it is a valid evaluation tool for activities of daily living [11] as well as cognitive screening [12]. For instance, researchers used to evaluate activities of daily living through a serious game platform by integrating game playing with a serious purpose. The aim of the study was to assess functional independence in older people through five tasks using the *Smart Aging* game [13]. In another study, the authors developed a serious game based on virtual reality to estimate cognitive disabilities during navigational tasks. Users had to follow arrows to get to special places and find the way back home on their own [14]. Even though the number of researches in this area has been significantly expanding, few researchers have addressed the development of IA systems with possible application into age-related MCI care based on the fact that users' daily cognitive abilities could be changed.

The aim of our study is to suggest an IA system for elderly people with MCI, and improve their ability to complete everyday tasks on their own. In our previous study [15], we implemented a task prompting system based on AR messages without any adaptive decision-making engine to evaluate the user's general cognitive condition. To increase the system's intelligence for making a correct reminder, we take advantage of an AR-based serious game assessment tool.

The suggested AR-based serious game in this work is potentially capable of providing entertainment and mental state examination simultaneously. Our target population and primary end-users are individuals experiencing MCI, who can understand the technology they are using and its expected assistance and risks. This preliminary study is focused on assessing the proposed system's usability by the older adults (with and without MCI) and considering their feedback.

In more detail, based upon the above suggestions, this paper proposes a system the following features:

- To implement the Internet of Things (IoT) system, the system benefits from the Message Queuing Telemetry Transport (MQTT) connectivity protocol for communication. All data values, such as indoor positioning data, embedded sensors, and cognitive serious game

  scores, can be published to the server in the system's automatic mode. Family or caregivers can also subscribe to each particular topic and override the message for the user to send reminders if needed.
- In the automatic mode, once an event happens in a specific indoor location, all the relevant fuzzy rules are checked on the server. Thus, following the decision-making algorithm, data values are updated. Both the user and the caregiver could receive appropriate notifications.
- To monitor the user's real-time location, the user wears an indoor localization tag. The localization tag sends the positioning information to the monitor module, and the monitor module receives all the published data.
- According to the AR serious game score, the performing mode of the IA system changes to the manual mode by real-time assessment of the user's cognitive state. This approach leads to saving the devices' energy, enhancing accuracy and performance.

## 2. Materials and methods

### 2.1. System architecture

As the daily routine of an elderly with MCI is contingent on their surrounding objects in the environment, environment conditions such as temperature, humidity, and $CO_2$ level is essential for designing a smart home environment. In our work, such data is published on the cloud layer by placing different sensors and actuators in several locations such as kitchen, bedroom and washing room.

Furthermore, Wireless Sensor Networks (WSNs) have been deployed in the local fog layer to collect data about the user's location. Three anchors are placed in the home environment to estimate the user's real-time location.

All the communication to our IA system is done over MQTT with data serialized in JavaScript Object Notation format (JSON). To avoid unnecessary transmission of the data, the performing mode of the IA system can be changed to the semi-automated mode based on the decision-making algorithm results.

We propose our model based on the localization tag worn by the user for positioning system. The application for making interaction with the end-user and assessing their cognitive functions is developed on an Android operating system, implemented on a smartphone.

In an attempt to send all the data to the cloud, we consider MQTT protocol as the messaging protocol. MQTT is a commonly used application layer protocol to transmit data between the devices in IoT architecture due to its simplicity and scalability [16]. We have utilized HiveMQ server as a MQTT message broker and Wemos D1 Mini modules for each sensor, actuator, and data coordinator in the positioning system to collect user's indoor data. Fig. 1 shows the general architecture of the proposed IA system in which the IA system works in its automated mode.

We have also designed a service-based application (based on the C# cloud service) for caregivers to present the data related to the events detected by the sensors and to control the actuators, embedded in the smart home environment [17]. The user cognitive functions state and game results can also be monitored by the caregivers.

### 2.2. Adaptive decision-making process

In this section, we consider an adaptive fuzzy method to convert our system into an intelligent and context-aware system that promotes monitoring of users and improving their ability to perform their daily tasks. To interact with the user correctly, the decision-making process should be adaptive and precise to avoid false notifications or false-positive AR messages. In some situations, such as completing daily tasks, people occasionally require assistance for memory recall.

In order to update each variable of the IA system on the cloud and making the right decision, we employed an adaptive fuzzy logic model that selects the most proper values. The model of the fuzzy adaptive decision-making is shown in Fig. 2.

A fuzzy controller has four main components: rule base, inference system, fuzzifier interface, and defuzzifier interface. In our system, we have designed two knowledge bases elicited from the user's cognitive state and the AR game score. These rule-bases are necessary for deciding when to provide aids and predict the user's target accurately. Now we go through each part of them in our model.

#### 2.2.1. Variables and membership functions

In our proposed model, the input variables are several types of data from embedded devices, the user's real-time location, and their cognitive state. The output variables are different types of AR messages which are received by the user or their caregivers. Table 1 shows input and output parameters, data types, and their membership functions in our work.

We can add other home objects and calculate the distance as a fuzzy input variable. By considering each input and output variable, membership functions are defined based on the data types and fuzzy rules.

Fig. 3 shows the membership function for the user's movement time as well as the AR game score. Output variables are voice, image, and text messages, defined as fuzzy singleton membership functions. Each message has an identification number (ID) and can be activated by particular inputs. A triangular membership function describes the states of the relay actuators.

#### 2.2.2. Fuzzy rule-bases

After defining each variable's membership function, we must build the rule-base, including all of the expert IF-THEN rules. In our IA system, two main rule-bases are essential for creating an adaptive decision-making process and multi-level support. The first one helps the user complete the activities independently in their daily life by showing AR messages or generating automatic changes such as actuators activation. The second one allows manual changes after real-time assessment of the user's cognitive state according to the AR-based serious game score. In this situation, some of the smart home sensors and reminders can be turned off or disabled.

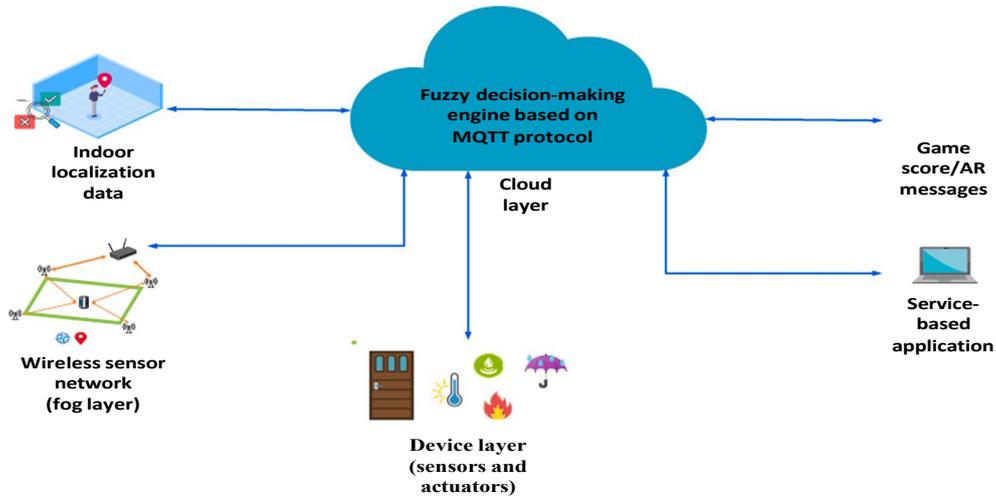

**Fig. 1.** Intelligent Assistive system architecture. All IoT data values, such as indoor positioning data, embedded sensors, and serious game scores, are published to the server. Following the fuzzy decision-making algorithm and serious game score, data values are updated in the automatic mode.

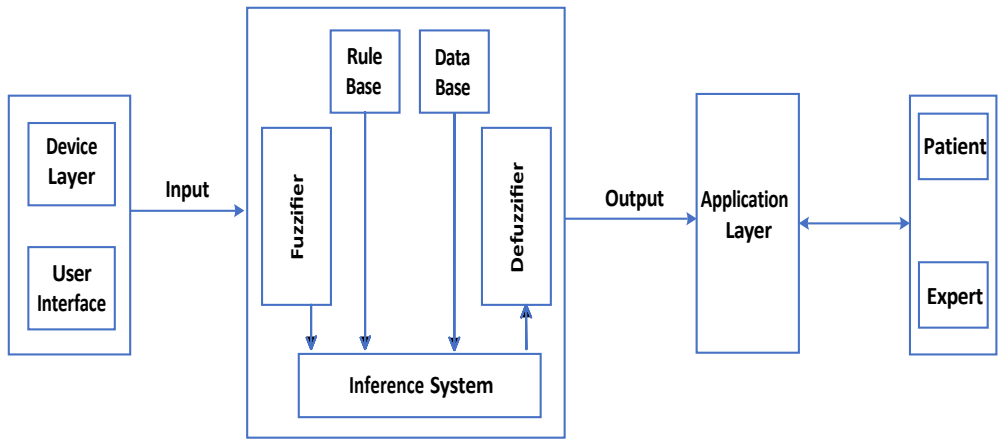

**Fig. 2.** Fuzzy adaptive decision-making model. Two knowledge bases are obtained from the user's cognitive state and the serious game score. User and caregiver receive appropriate notifications based on the fuzzy rules' outputs.

**Table 1**
Description of Inputs and outputs of the fuzzy system.

| Parameter | Fuzzy membership function | Data type |
| --- | --- | --- |
| Game score | Low, high | Linguistic |
| Game | Start, Stop | Boolean |
| Time | Early morning, morning, early afternoon, late afternoon, evening, night, midnight | Linguistic |
| Distance | Near, far, very far | Linguistic |
| Humidity | Very dry, dry, humid, very humid | Linguistic |
| Temperature | Very cold, cold, cool, mild, warm, hot, very hot | Linguistic |
| Reminder | Yes, No | Boolean |
| Movement | Low, medium, high | Linguistic |
| Flame detection | Yes, No | Boolean |
| Gas detection | Yes, No | Boolean |
| Relay status | Yes, No | Boolean |
| Voice message | 1, 2, … | Integer |
| Image message | 1, 2, … | Integer |

Once an event occurs (for example, medication time alarm), input data values such as location, AR game score, and sensors data value are published on the cloud via MQTT protocol. Moreover, fuzzy rules are checked, and corresponding output variables are updated and published on their topics. Then, according to which rule-bases are activated, the user receives different types of messages, or the user's manual changes can be enabled.

The adaptive fuzzy decision-making algorithm is capable of scaling up to more rules to improve the user's independence. The location of different objects and the most used ones can also be added to the database.

Therefore, new rules or new objects can be added, or the other rules can be removed and updated by the caregivers remotely based on the knowledge and dependent of each user's lifestyle, and they can manually start a scenario or send simple reminders to the user. The fuzzy rules must be endowed with a base of knowledge of each user provided by the caregiver, so all the activities are adapted to their preferences [18].

Some parts of such scenarios and their fuzzy rules are presented in Table 2. These scenarios are precisely defined to choose the most proper ones for the first experiment with individuals with cognitive impair- ments [4,18].

According to [4], elderly people with MCI often lose their ability to sequence in activities such as putting on clothes or cooking. When the user walks in different rooms and tries to interact with other objects, the IA system tries to predict the user's target based on the distance between the user and predefined objects and other inputs.

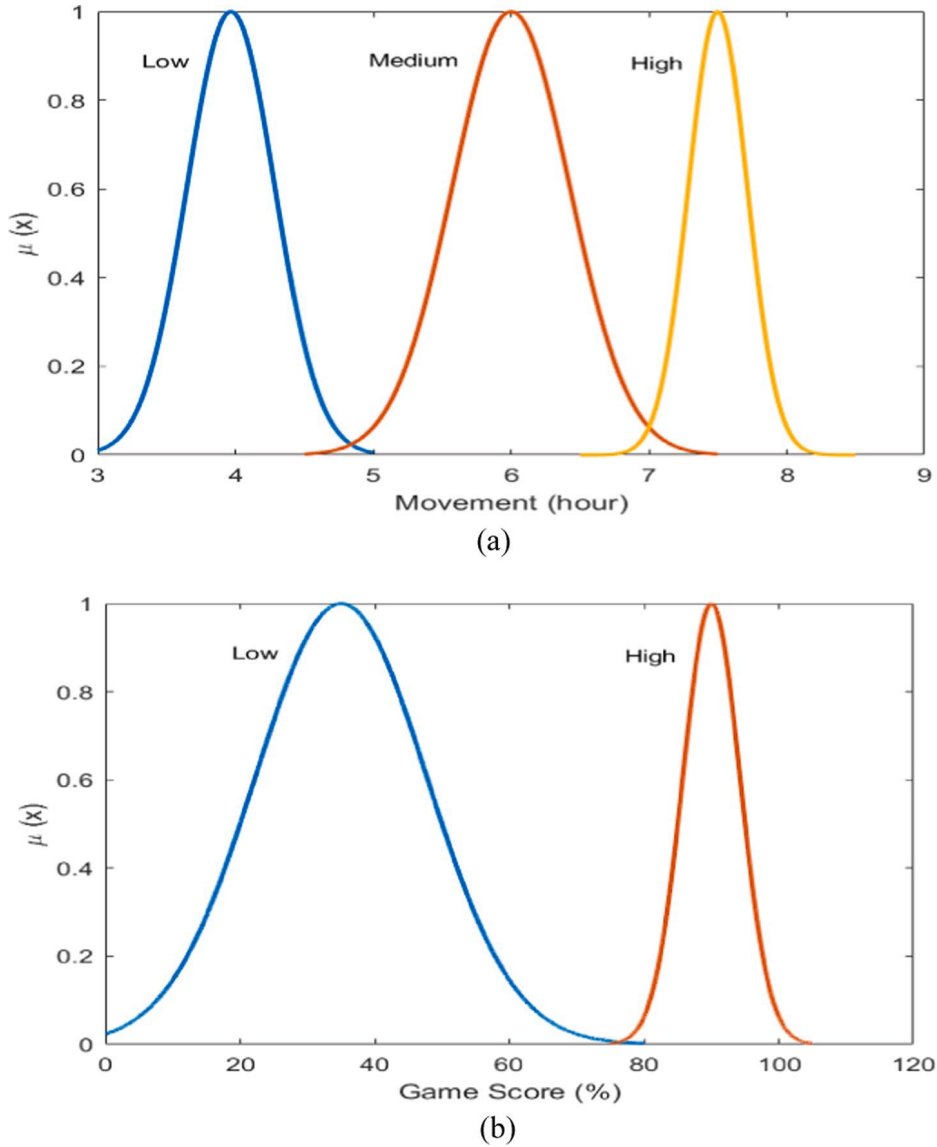

**Fig. 3.** Gaussian membership function of the user's (a) movement and (b) game score.

Lack of movement is also an important issue that should be considered in the users' daily life. The system measures the user's movement to check this factor by localization tag worn by the user. If he moves between 3.5 and 4.5 h per day, the IA system sends a reminder to the user and their caregiver [19]. In another scenario, once the user is near the drawer of a wardrobe or closet in the bedroom, the IA system can suggest to them what to wear. This approach can help the user make everyday decisions before they get nervous.

### 2.3. Real-time position estimation

To provide users' interaction with the objects and monitoring platform for caregivers, we need the real-time location of the user. There are many challenges introduced by GPS into the system such as high energy consumption and vulnerabilities that may cause poor accuracy rates in specific environments, for example, indoor locations. Based on the literature, the most accurate beacon-based local positioning system (LPS) solutions are those using ultrasound [20], or ultrawideband (UWB) radio signals [21].

However, ultrasound has a weak capability of a limited maximum range (about 10 m) and cannot pass through walls; thus, its coverage is limited by the number of beacons to install [22]. UWB can penetrate walls in buildings and is able to resolve individual multipath mechanisms due to its large bandwidth. Therefore, we have designed our positioning system based on the UWB, because user's location can be changed over time through different rooms. To achieve this goal, a new prototypical indoor monitoring method developed on a low-cost UWB WSNs, based on IEEE 802.15.4 and IEEE 802.15.4e standards, has been created as a part of our IA system [23].

It is mentioned that the location of the frequently used objects is predetermined; therefore, the distance from the objects can be simply estimated. Locations of the objects are supposed to be fixed and stored on the server. Thus, to calculate the user's distance from a predefined object, the Euclidean distance is easily assessed between the user and locations of the objects. A true-range multilateration technique is employed as a positioning algorithm in the implemented indoor positioning system [24].

Position calculation consists of two main parts; finding distances from the tag to the anchors, and solving the location inverse geometry problem. The WSNs based on UWB is implemented using a single-chip wireless transceiver, Decawave (Dublin, Ireland) Sensor DW1000 Module [22]. The impulse-based UWB facilitates a precise ranging and

**Table 2**
Examples of fuzzy rules.

| Event | User's location | Fuzzy rules If | Fuzzy rules Then | Command type |
|---|---|---|---|---|
| Entertainment/ cognitive assessment | Not specific | *Time* is "*evening*" | *Game* is "*start*" | Launching AR game |
| Medication schedule | Not specific | *Time* is "*morning*" *Gas detection* is "*Yes*" | *Voice message* is "*1*" and *Image message* is "*1*" | Reminder |
| Leaving the stove on | Not specific | | *Relay status* is "*Yes*" and *Voice message* is "*2*" | Alert |
| Trouble with cooking | Kitchen | *Distance\** is "*near*" | *Voice message* is "*3*" and *Image message* is "*2*" | Reminder |
| Putting on clothes | Bedroom | *Distance* is "*near*" | *Voice message* is "*4*" and *Image message* is "*3*" | Reminder |
| Meal time | Not specific | *Time* is "*early afternoon*" | *Voice message* is "*5*" and *Image message* is "*4*" | Reminder |
| Lack of movement | Not specific | *Time* is "*early afternoon*" | *Voice message* is "*6*" | Alert |
| High game score | Not specific | *Game score* is "*high*" | *Reminder* is "*No*" | Disable reminders |

*Distance between the user and particular object (refrigerator, drawer, etc.)

high-accurate localization of the network nodes [25,26].

Our range finding algorithm is a double-sided and two way ranging based on time difference of arrival (TDoA) value [27]. In the designed local fog, calculated ranges are always transmitted between the tag (processor: STM32F4 Cortex-M4) and the anchors (processor: STM32F1 Cortex-M3), sniffed by the data coordinator.

Furthermore, the data coordinator concentrates and encapsulates all the transferred distances on the local fog and publishes them in JSON format via MQTT protocol to the cloud server. For activity recognition, pedometer for estimating user's total movement and fall detection, all the motion processing libraries available in X-CUBE-MEMS1 and can be simply implemented.

*2.4. Visualizations*

The IA system uses two different user interfaces (UI) to interact with the users and their family. Both user interfaces are developed by uti- lizing ARCore SDK (Software Development Kit) for Unity and Android features [28]. In this section, we consider the applications and their features for making interaction with the user, monitoring service for the caregivers, and the development process of designing a serious game environment.

*2.4.1. AR interaction*

Unity is a cross-platform game engine that can be applied to design games and simulations for computers, consoles, and smartphones [29]. To display the user's localization information clearly, a 3D indoor positioning system based on the Unity 3D platform is created. This configurable user interface can represent complex indoor area geographic information and enhance the user experience [30].

Furthermore, to monitor and track the user's real-time location and provide their interaction with the objects, the user wears an indoor localization tag. For presenting and transmitting positioning data in JSON format via MQTT protocol to the cloud server, MQTTnet and Json. NET libraries were also employed in the Unity application. The localization tag wearing by the user sends position information to the monitor module, and the monitor module receives all the positioning data.

All the data related to the localization system and AR messages can be stored on the cloud for further studies. To improve the user's safety, caregivers can add the location of their home's danger zone. Hence, the IA system can provide warnings if the user is near to these locations. The second user interface is built on Android Studio and SDK tools to interact with the individual with MCI as a personal assistant device [31]. Android Studio is the official integrated development environment (IDE) for the Android operating system created particularly for Android development. In this work, two different AR messages are defined to send reminders or alerts to the user, including visual and audio messages.

The AR messages show information using virtual data based on the predefined fuzzy rules. These multimedia contents are for taking medication, playing AR cognitive training games, reminding meal times, and alerting danger zones. For example, to recall medication schedule, a voice message, "Let's take our medicine," and an image message indi- cating the drug's picture are defined. These messages must not remind users' disability.

Based on the similar studies and their recommendations on designing serious games for people with MCI [32–34], we have also considered other designing factors in the AR interaction, including:

. A summary screen showing scores, errors or are included in the game time to foster competition and make an appropriate interaction.
. Instructions before the game about how to solve game tasks are kept brief and concise.
. Only one active screen area to interact with is designed.
. Large and highly contrasted icons are used.
. The game is played from the user's perspective to prevent confusion with avatars.
. The difficulty level of the game increases when the user completes the first task correctly, so this can result in early success to foster motivation and prevent frustration.

*2.4.2. AR-based serious game and scenarios*

Initiating treatment at the early stage of cognitive problems would slow the progression of the situation and expand the individual's quality of life [35]. In this study, an AR-based serious game provides enter- tainment and mental state examination simultaneously. The IA system suggests playing the game several times per day to get the user involved and keep them active. Some parts of the game's features are as follows:

. It is designed based on "World Tracking" that allows users to put an AR object anywhere they would like in the camera view.
. Before starting each task, the system provides some details within the questions to trigger memory through voice interactions.
. After performing each task, users' score is calculated based on the number of right and wrong decisions automatically.
. The time duration of each task and total time response are measured automatically.

The AR serious game consists of a simulation of daily living situations with five tasks: finding a particular object, remembering colors, and arranging numbers. These tasks are defined to evaluate different cognitive functions such as pattern separation and completion, visuo- spatial and episodic memory, decision-making ability, concentration, and overall processing speed by measuring response time [33].

Performing such tasks may also result in improving the user's cognitive abilities ultimately. Furthermore, based on the fuzzy rules, the tasks can be suggested to the user to perform several times per day for assessing their cognitive status and rehabilitation program. To prevent reappearance, we have designed three series of the game scenarios in which each task contains different 3D objects. In every session, the IA system randomly selects one of the game series for recommending to the user. Table 3 describes the tasks and outputs measures.

Before beginning the assessment process and game scenarios, the user should look at the whole objects for 10 s (second) to memorize their location and color, as shown in Fig. 4 (a). Audio messages are provided before starting each task to help the user realize the task requirements, for example, which object's color should be recalled [33]. Fig. 4 illus- trates the game steps based on the tasks presented in Table 3.

In more detail, Fig. 5 (a) shows the task 4 in which the user should identify the unnatural placement of objects, for instance, headphones in a sink [33]. In the next step, the user observes a series of numbers on the screen for 5 s, and after that, he must choose the correct sequence of numbers in the picture frames. This step is proposed based on the fact that people with mild cognitive impairments usually lose their ability to sequence [4]. Fig. 5 (b) demonstrates this step of the game. Finally, the user's final score according to the total time response and the number of true (T) and false (F) answers are calculated.

The total response time must not be more than 10 min, which is the normal cognitive assessment test duration [34]. Fig. 5 (c) indicates an example of the assessment result that is sent to the fuzzy decision- making system for evaluating the level of support, this result is not included in the statistical analysis.

The IA system is designed based on error-less learning to encourage the user to recall different events. If the user gives more correct answer, the system vanishes reminders and turns off some smart home sensors. Thus, it can improve the user's self-management and self-care, and it would also slow the progression of the disease [3].

To make a user-friendly interaction, the game is presented on a touch screen for performing different tasks, because most of the older adults can interact with the touch screen without any training [36]. We have used different 3D objects for creating an AR environment, and an event- based creation supporting unit for defining interaction between the virtual space object in a format of event script [37].

To build the AR serious game, different virtual objects are incorpo- rated into the web-based AR platform, ZapWorks Studio and ZapWorks Designer (Zappar Ltd., United Kingdom) [38]. The platform allows us to upload multimedia contents to create an AR experience that lets users view the virtual objects through the mobile device application, Zappar. This application could be efficiently run on both Android and iOS operating systems. In the current study, we have used a Samsung Galaxy S9 Plus smartphone, with an Octa-core processor, 6 GB RAM, and dual 12MP rear camera to view the AR experience, with each virtual object being displayed on the screen.

**Table 3**
Descriptions of tasks and outcomes measures.

| Tasks | Outcome measures |
|---|---|
| Task 1: retrieving objects location | Time on the task 1 |
| | Task completion: selecting the correct location |
| | Score: 1 point for the correct answer |
| Task 2: memorizing objects color | Time on the task 2 |
| | Task completion: selecting the correct color |
| | Score: 1 point for the correct answer |
| Task 3: recognizing an extra object | Time on the task 3 |
| | Task completion: selecting the correct object |
| | Score: 1 point for the correct or wrong answer |
| Task 4: identifying unnatural placement of objects | Time on the task 4 |
| | Task completion: selecting the correct picture |
| | Score: 1 point for the correct or wrong answer |
| Task 5: recalling sequence of numbers | Time on the task 5 |
| | Task completion: selecting the correct numbers sequentially |
| | Score: 1 point for the correct answer |

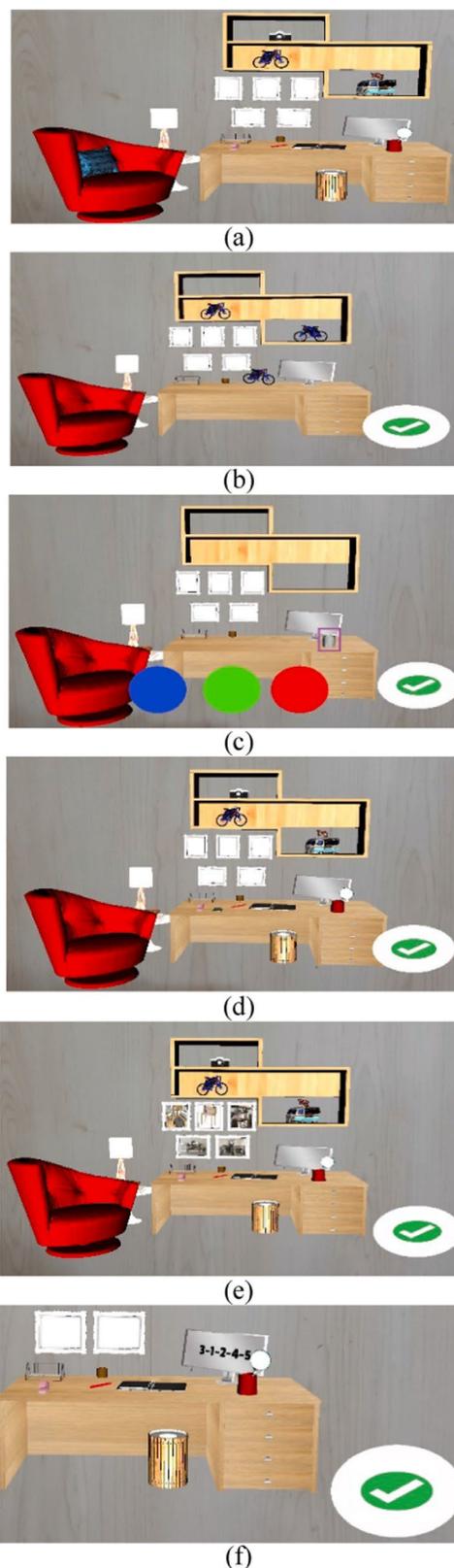

**Fig. 4.** Sc... the main environmen... orizing the object col... identifying unnatural placement of the objects in the picture frames, and (f) Task 5: recalling sequence of the numbers on the screen.

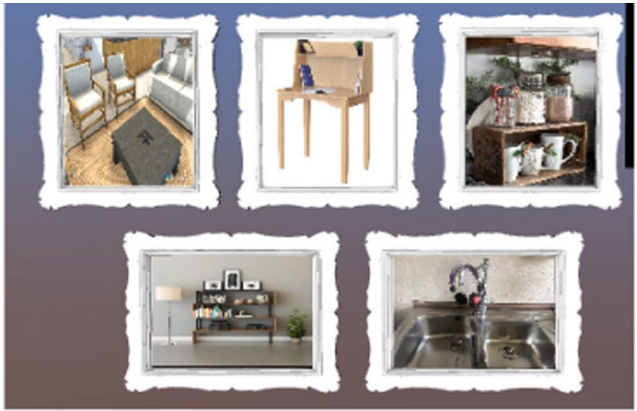

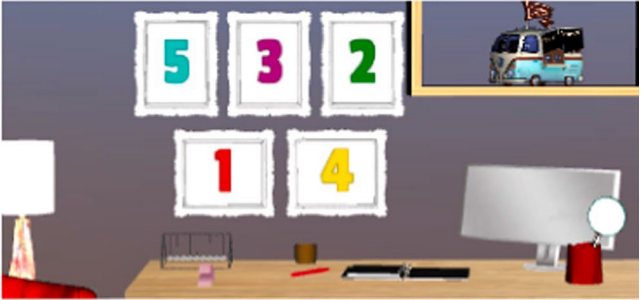

**Fig. 5.** Illustration of response of a user (a) identify the unnatural placement of objects, (b) choose the correct sequence of numbers in the picture frames, and (c) final score is sent to the decision-making system.

*2.5. Study participants*

We set out to validate the IA system and the serious game in a group of individuals aged above 50 years to be age-matched with the end-users. They comprised 37 older adults, including 16 females and 21 males. All participants were volunteers recruited from universities and social media. They all lived independently and had active social and cognitive lives. Persian was the mother tongue for all participants in the study. Before playing the AR serious game, participants underwent the MoCA [39]. The MoCA is a 30-point test, including a set of cognitive subtests, concerning six different cognitive domains.

Based on the literature, a cutoff of 26 (scores of 25 or below indicate impairment) yields the best balance between sensitivity and specificity for the MCI and other groups [40]. Thus, in our study, we have divided participants into two groups based on their MoCA results: (A) control group (MoCA ≥ 26) and (B) MCI group (MoCA < 26). Table 4 shows the number of people from each group and their gender.

**Table 4**
Group characteristic in the study (A: control group, B: MCI group).

| Participants (n = 37) | Total number | Female | MoCA score |
|---|---|---|---|
| Group A | 26 | 12 | ≥26 |
| Group B | 11 | 4 | <26 |

All participants first filled a questionnaire including demographic information, personal and medical history. They then completed a training in the game in order to get familiarize with the interactions of virtual objects for final evaluation. Table 5 describes all the participants' demographics in details.

*2.6. Statistical analysis*

We have analyzed all the data and game scores to consider serious game accuracy compared to MoCA scores. We chose the MoCA as our reference screening tool as it verified to be sensitive to recognize MCI and it is an accurate screening measurement of cognitive ability. The significance level was set $\alpha < 0.05$ for all analyses. The SPSS 26.0 statistical software package was used to perform all the statistical analyses. The data was normal distribution and homogeneous, so Pearson correlation coefficient was used for categorical variables.

### 3. Results and discussions

In this section, we present the technical feasibility and usability of the system via simulations and user experiences. First, the IA system's accuracy in detecting the danger zones and making appropriate recommendations have been assessed. Then, the serious game results with 37 participants compared with their MoCA test scores have been provided. Finally, we evaluated the system's performance in working in different modes.

*3.1. Forbidden zones*

In this experience, we evaluated the system's accuracy in detecting danger zones area. Four different danger areas have been defined with equal sizes of $0.52 \times 0.7 \times 0.23$ m$^3$ in an $8.5 \times 4.6 \times 2$ m$^3$ room space, as shown in Fig. 6. Fig. 7 illustrated these zones, which were determined in the Unity game engine to display the user's localization information clearly.

In a typical scenario, once the user entered a dangerous zone (for example, the fireplace), they received an image message alarm. This alarm reminded them to keep away from the area. The near membership function was defined from 1 to 5 dm (the danger zone). Fig. 8 demonstrates the AR message showing the fire alert sent to the user by activating the fuzzy rule number 16. Other reminders could be for pharmacological management and completing daily tasks.

The positioning data and the movement time of the user was recorded during the experiment. According to the system's operating mode and the users' cognitive state, if the users' movement duration was lower than 4.5 h, they received a voice message. This audio message (ID: 13), which was sent by the IA system, reminded the users their lack of movement [19]. If the movement duration was higher than 4.5 or the user's serious game result was high, they did not receive any messages or alarms. Table 6 summarizes the results of these experiments in different conditions based on the data values.

**Table 5**
Participants' demographics.

| Participants | Average | Minimum | Maximum | Standard deviation |
|---|---|---|---|---|
| Age of group A | 60.07 | 51 | 81 | 7.98 |
| Age of group B | 65.73 | 56 | 80 | 7.91 |
| MoCA score of group A | 27.71 | 26 | 29 | 1.069 |
| MoCA score of group B | 23.36 | 21 | 25 | 1.36 |

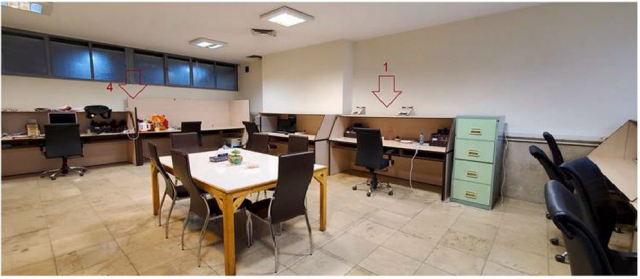

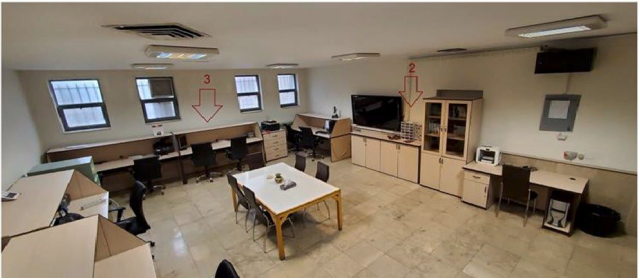

**Fig. 6.** Four danger zone areas are defined for evaluation of the localization system in the experimental condition.

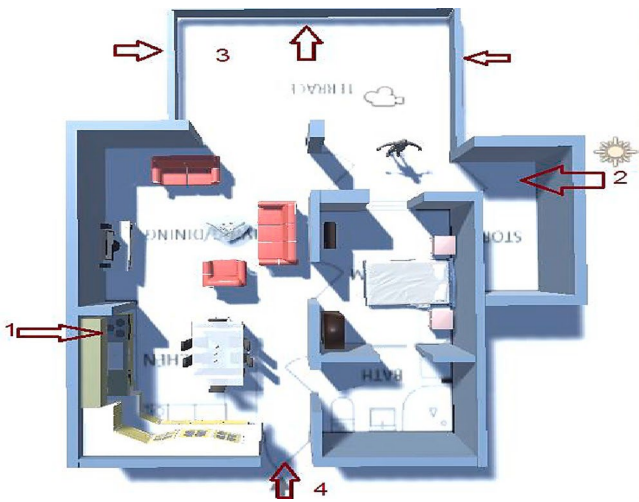

**Fig. 7.** Model of the 3D-indoor positioning system and four danger zone areas.

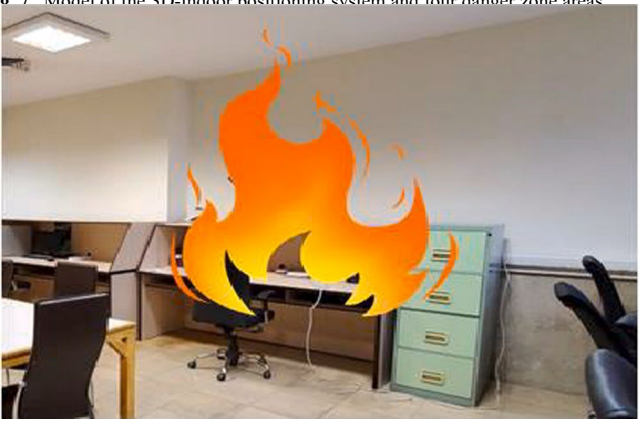

**Fig. 8.** Sample of an image message received by the user when a specific fuzzy rule was activated, picture message number 16 (fire alert).

### 3.2. Serious game

We have evaluated the serious game's accuracy with 37 participants and compared the game score with the users' MoCA test scores. Table 7 compares both scores in group A and group B.

We used the Pearson correlation coefficient for control group participants with respect to the parametricity of the data and applied the Spearman correlation coefficient due to the nonparametric data for the MCI group. These correlations have been calculated between the score of the MoCA and the score of the game, and the total time and participants' age. The results are given in Table 8 and Table 9.

According to the results, the total score of MoCA had a significant correlation (0.911) with the total score of the game in the control group, while the correlation coefficient of Spearman in the MCI group has a lower correlation (0.722). Furthermore, there was no correlation between the total score of MoCA and the total time of the game in the control group. While in the MCI Group, there was a relative correlation (0.773) between the total score of MoCA and the total game time duration. This may indicate that the response time in the MCI group is a more effective factor than the control group.

Also, in both groups, there was no correlation between the score of the game and the age of the participants, which means that the age of the participants was not a determining factor for the game score.

Table 10 compares the scores of MoCA and serious game results. These grades showed that the participants with the MoCA score higher than 27, could get at least 4 scores in the serious game.

Spearman correlation coefficient between the total score of MoCA with total time was 0.692 for the participants who scored 3 points in the game. This number increased to 0.841 for the participants who scored 2 points and 1 point in the game. However, a comprehensive evaluation cannot be concluded from these correlations due to the very small amount of data for each group of participants who received the same score of 1 to 3 from the game.

We also performed a $T$-test for the total score of the game to determine a meaningful performance difference between the control and MCI group members. The results of this evaluation showed that the MCI group had a lower overall game score (2.09 ± 0.893) compared to the control group participants (4.21 ± 0.70); $t(37) = 6.939$, $p < 0.001$.

Also, the result of the $T$-test for the total game duration time between the control and MCI groups indicated that the MCI group had a higher total game time duration (5.13 ± 2.03) than the control group (3.35 ± 1.9); $t(37) = -2.275$, $p < 0.05$.

This means that there is a considerable difference between the control group and MCI group participants' cognitive performance. Our evaluation findings were in line with previous results indicating that the touchscreen was well accepted by participants without computer experience based on the participants' feedback [41,42].

### 3.3. System operation modes

We considered a wide range of devices that must cover the maximum

**Table 6**
Result of the experimental test.

| Movement (hour) | Serious game result (out of 100) | Voice message ID | Danger zone number | Image message ID |
|---|---|---|---|---|
| 1.15 | 55 | 13 | 1 | 16 |
| 5.30 | 48 | – | 2 | 17 |
| 5.25 | 97 | – | 3 | 18 |
| 3.40 | 76 | 13 | 4 | 19 |

**Table 7**

Comparison between group A and group B game results.

| Participants | Average | Minimum | Maximum | Standard deviation |
|---|---|---|---|---|
| Serious game score (out of 5) group A | 4.21 | 3 | 5 | 0.70 |
| Serious game score (out of 5) group B | 2.09 | 1 | 3 | 0.83 |
| Serious game time duration (mins) group A | 3.35 | 1.38 | 7.47 | 1.9 |
| Serious game time duration (mins) group B | 5.13 | 2.42 | 8.83 | 2.03 |

**Table 8**

Pearson correlation coefficient between MoCA scores and total serious game score, total response time and age in the control group.

| Pearson Correlation in group A | Total Serious Game Score | Total Response Time | Participant Age |
|---|---|---|---|
| MoCA Total Score | 0.911 | -0.343 | 0.003 |
| Total Serious Game Score | 1 | -0.115 | 0.031 |

**Table 9**

Spearman Rank Correlation (Spearman's Rho) between MoCA scores and total serious game score, total response time and age in MCI group.

| Pearson Correlation in group A | Total Serious Game Score | Total Response Time | Participant Age |
|---|---|---|---|
| MoCA Total Score | 0.722 | -0.773 | -0.278 |
| Total Serious Game Score | 1 | -0.477 | 0.017 |

**Table 10**

Comparison between the scores of MoCA and serious game results.

| Group | Total MoCA score | Total serious game score |
|---|---|---|
| Control Group | 29≤ | 5 |
| MCI Group | 28 | 4–5 |
| | 27 | 4 |
| | 26 | 3 |
| | 25 | 3 |
| | 24 | 2–3 |
| | 23 | 2 |
| | 22 | 1–2 |
| | 21 | 1–2 |

area of the user's environment. The number of devices ranged from 10 to 15 to realize different workloads, including various sensors and actuators as we mentioned in Section 2.1. Specifically, for each number of devices, we simulated the two different cases based on the system's operation mode. The complexity of the decision-making process decreased from automated mode to semi-automated mode, as the number of active devices declined from 22 to 14.

In the functionality test, we consider the number of audio messages sent to the user, and the number of IoT devices enabled during the experiment. Fig. 9 illustrates the numbers relevant to different types of AR messages included in each case and received by the user. As shown in Fig. 9, the number of alarms and reminder messages decreased as the user's game score increased.

The results indicated that the user required lower reminders while their cognitive test score was higher than previous session. These also led to less data loss in the decision-making algorithm and during the process of publishing or subscribing to the messages. Moreover, the operation in semi-automated mode can cause less need for battery recharging or replacement [43].

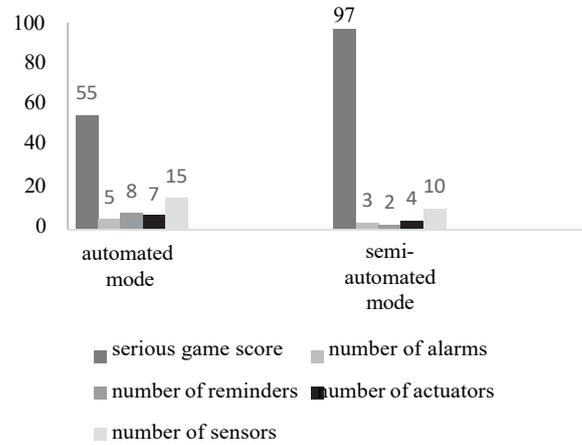

**Fig. 9.** Workload scenarios description. The number of alarms and reminder messages decreased as the user's game score increased.

### 3.4. Limitations

The target population and primary end-users of our proposed system are older adults experiencing cognitive impairments who can understand the technology they are using and its expected assistance and risk. This study was focused on assessing the proposed system's usability by the users without cognitive impairments and considering their feedback. However, this study has some limitations.

The space validity of this work was limited, for part of the study was presented in a laboratory environment. The IA system prototype can be evaluated with more users, and it includes more scenarios and functionalities. Therefore, although the scenarios technical requirements are taken into account in our work, the particular content of the messages must be customized for each user before the system's implementation. To achieve this goal, the IA system is capable of adding more fuzzy rules and being customized for each person.

### 4. Conclusions and future work

Memory loss and decision-making difficulties are the main symptoms of MCI experienced by elderly people. The person forgets recent incidents and have problems with remembering the information. Such situations have a considerable effect on the confidence and quality of life of both the elderly people and their caregivers. The aim of this study was to introduce an IA system for individuals with age-related MCI, and improve their ability to complete everyday tasks on their own without compromising their privacy. We designed, implemented, and evaluated an IA system based on the AR serious game and IoT to help users make everyday decisions more independently through interaction and recall memory of events.

Several ready-to-use libraries were used to facilitate system implementation on smartphones and computers. We evaluated the system's technical feasibility and usability via simulations and users' experiences with 37 elderly participants. The result of our evaluation implied that the total score of the serious game had a high correlation with the total score of MoCA in the control group, and there was an acceptable correlation in the MCI group. Also, our game was capable of showing a considerable difference between the two groups of control and MCI participants' cognitive performance. This indicates the high accuracy of the system in estimating users' cognitive states compared to the traditional assessment tools.

Moreover, the touchscreen interaction and verbal reminders created a natural way of interaction and were well accepted by the participants without computer experiences. The system response in the semi-automated mode also caused less data loss than the automated mode, as the number of active devices decreased.

The findings of this study support the idea that an IA system can potentially help older adults in the home environment and can be evaluated by users with cognitive impairments according to participants' feedback. Serious games also have a positive impact on presenting the possibility of reconstructing a functional environment for the users. Some future plans could be the study of historical data, including daily serious game scores, to evaluate the user's mental and physical health condition.

**Declaration of Competing Interest**

The authors declare that they have no known competing financial interests or personal relationships that could have appeared to influence the work reported in this paper.

**Acknowledgments**

The authors gratefully acknowledge the families and the individuals that participated in this study.